\documentclass[twocolumn]{aastex62}
\pdfoutput=1 
\usepackage{amsmath,amstext}
\usepackage[T1]{fontenc}
\usepackage{apjfonts} 
\usepackage[figure,figure*]{hypcap}

\usepackage{url}

\usepackage{hyperref}
\usepackage[hyphenbreaks]{breakurl}

\def\simpropto{\lower.2ex\hbox{$\; \buildrel \sim    \over \propto \;$}}

\def\fun#1#2{\lower0.837ex\vbox{\baselineskip0ex\lineskip0.209ex
  \ialign{$\mathsurround=0ex#1\hfil##\hfil$\crcr#2\crcr\sim\crcr}}}

\def\sles{\lower2pt\hbox{$\buildrel {\scriptstyle <}
   \over {\scriptstyle\sim}$}}

\def\sgreat{\lower2pt\hbox{$\buildrel {\scriptstyle >}
   \over {\scriptstyle\sim}$}}

\begin{document}

\title{The Superoutburst Duration versus 
 Orbital Period Relation for AM CVn stars}
  
\shortauthors{Cannizzo \& Ramsay}
\author{J. K. Cannizzo}
\altaffiliation{Deceased 30 December 2018}
\affiliation{CRESST/Joint Center for Astrophysics,
                 University of Maryland, Baltimore County, Baltimore, MD 21250, USA}
\affiliation{Astroparticle Physics Laboratory, 
  NASA-Goddard Space Flight Center, Greenbelt, MD 20771}
\author{G. Ramsay}
\affiliation{Armagh Observatory,  College Hill,  Armagh BT61 9DG, U.K., gavin.ramsay@armagh.ac.uk}
\correspondingauthor{Gavin Ramsay}
\email{gavin.ramsay@armagh.ac.uk}
                   
\begin{abstract}
We examine the relationship between superoutburst duration $t_{\rm
  dur}$ and orbital period $P_{\rm orb}$ in AM CVn ultra-compact
binary systems.  We show that the previously determined steep relation
derived by \citet{Levitan2015} was strongly influenced by the
inclusion of upper limits for systems with a relatively long orbital
period in their fit.  Excluding the upper limit values and including
$t_{\rm dur}$ values for three systems at long $P_{\rm orb}$ which
were not considered previously, then $d \log (t_{\rm dur})/ d \log
(P_{\rm orb})$ is flat as predicted by \citet{CannizzoNelemans2015}
\end{abstract}


\section{Introduction}

AM CVn binaries are interacting binaries with orbital periods $P_{\rm
  orbital}$ between about 5 and 65 min in which a Roche-lobe filling
low-mass white dwarf (WD) secondary transfers matter to a more massive
WD primary (\citealt{Paczynski1967}, \citealt{Faulkner1972},
\citealt{Solheim2010}, \citealt{Nelemans2015}). Since the mass losing
star is degenerate and has an inverse mass-radius relation, mass-loss
drives the binary to longer orbital periods.  Thus AM CVn systems are
thought to have evolved beyond the minimum orbital period that
separates degenerate and non-degenerate secondaries (see
\citet{Littlefair2008}).  AM CVn's make up one subclass of the
cataclysmic variables (CVs).

Due to the faintness of these systems, it has taken decades of work
just to identify more than a handful of systems and to begin to
characterize their long term light curves. The last few years have
witnessed a break-through in terms of significantly adding systems
with orbital period determinations and long term light curves so as to
quantify outburst behavior. Dedicated transient surveys have
facilitated this advance. \citet{Ramsay2012} presented results of a
$\sim$2.5 yr monitoring program of 16 AM CVn systems using the
Liverpool Telescope, and showed the existence of an `instability
strip' in the orbital period distribution ranging from $\sim$25 to 45
min in which outbursts occur, consistent with earlier theoretical
predictions (\citealt{Smak1983a}, \citealt{TsugawaOsaki1997}). Surveys
such as the Palomar Transient Factory (PTF)
\citep{Levitan2013,Levitan2015} have been very successful in
discovering new AM CVn systems and characterizing their long term
variability properties.

The accretion disk limit cycle (ADLC) model was developed to account
for hydrogen rich dwarf nova (DN) outbursts (\citealt{MeyerMeyer1981},
\citealt{Smak1984}). It proposes that material accumulates in the
accretion disk during quiescence and accretes on the WD during
outburst when a certain critical surface density is attained.  There
exists a critical rate of accretion ${\dot M}_{\rm crit}$ above which
disks are predicted to be stable (the novalikes - NLs) and below which
outbursts occur (DN). The Z Cam (ZC) systems with mass transfer rate,
${\dot M}_T\simeq {\dot M}_{\rm crit}$ could show either behavior,
given that ${\dot M}_T$ is seen to vary on long time scales.
\citet{Smak1983b} and \citet{Warner1987} investigated the relationship
between ${\dot M}_T$, orbital period, disk radii and sub-class of
dwarf novae and concluded the ADLC could account for NL and ZC
systems. This was reinforced by \citet{Dubus2018} who used a sample of
$\sim$130 CVs with parallaxes in the {\sl Gaia} DR2 sample to test the
predictions of the ADLC model.

Short orbital period DNe show two kinds of outburst - short or normal
outbursts in which the system quickly rises to maximum light and then
returns to quiescence, and long or superoutbursts which are of longer
duration and higher amplitude than short outbursts. In long outbursts
there is a time interval of slow decay following maximum light and
then a rapid return to quiescence (e.g. \citealt{Cannizzo2002}).
Superoutbursts are defined by the presence of `superhumps',
modulations in the optical light at periods slightly greater (positive
superhumps) or slightly less than (negative superhumps) the orbital
period.

Like the hydrogen accreting CVs, the AM CVn systems also show normal
and superoutbursts. The normal outbursts last only a few days and its
only recently with high cadence surveys like PTF that systems such as
PTF1J0719+4858 (an AM CVn binary with an orbital period of 26.8 min)
have shown evidence for normal and superoutbursts
\citep{Levitan2011}. Given the difficulty of identifying normal
outbursts from AM CVn's most of the outbursts observed from these
binaries are likely to have been superoutbursts.

\citet{CannizzoNelemans2015} applied the ADLC model to the instability
strip found by \citet{Levitan2015} for the AM CVn systems, using ADLC
scalings from \citet{Kotko2012}, and made inferences for three
scalings of interest: (i) the mass transfer rate ${\dot M}_T$ versus
$P_{\rm orb}$, (ii) the recurrence time $t_{\rm recur}$ for outbursts
versus $P_{\rm orb}$, and (iii) the superoutburst duration $t_{\rm
  dur}$ versus $P_{\rm orb}$.  If one assumes a power law ${\dot M}_T
= A P_{\rm orb}^n$, then based on the observed width of the strip (and
assuming $m_1 = 0.6$ for the primary), one finds (i) ${\dot
  M}_T\propto {P_{\rm orb}}^{-5.2}$, (ii) $t_{\rm recur} \propto
{P_{\rm orb}}^{7.4}$, and (iii) $t_{\rm dur} \propto {P_{\rm
    orb}}^{0.4}$.  These derive from simple considerations of the ADLC
model.  The steep inverse mass transfer rate ${\dot M}_T (P_{\rm
  orb})$ is consistent with stellar evolution models, and the steep
$t_{\rm recur} ({P_{\rm orb}}) $ relation is consistent with that
found by \citet{Levitan2015}.  However, the relatively flat $t_{\rm
  dur} ({P_{\rm orb}})$ scaling is inconsistent with the much steeper
relation $t_{\rm dur} \propto {P_{\rm orb}}^{4.5}$ found by
\citet{Levitan2015}.  In this work we investigate this discrepancy.

\section{Accretion Disk Physics}

In the accretion disk limit cycle model, gas accumulates in quiescence
and accretes onto the central object in outburst (e.g.,
\citealt{Cannizzo93}, \citealt{Lasota2001} for reviews).  The phases
of quiescence and outburst are mediated by the action of heating and
cooling fronts that traverse the disk and bring about phase
transitions between low and high states, consisting of neutral and
ionized gas, respectively.  During quiescence, when the surface
density $\Sigma(r)$ at some radius within the disk exceeds a critical
value $\Sigma_{\rm max}(r)$, a transition to the high state is
initiated; during outburst, when $\Sigma(r)$ drops below a different
critical value $\Sigma_{\rm min}(r)$, a transition to the low state is
initiated.  Low-to-high transitions can begin at any radius, whereas
high-to-low transitions begin at the outer disk edge.  This situation
comes about because in the outburst disk $\Sigma(r) \propto r^{-3/4}$
(roughly) and the critical surface densities both increase with
radius.  Since the disk mass accumulated in quiescence is bounded by
$\Sigma_{\rm max}(r)$ and $\Sigma_{\rm min}(r)$, one can define a
maximum disk mass
\begin{equation}
 M_{\rm disk,  max} = \int 2\pi r dr \Sigma_{\rm max}(r)
\end{equation}
  and a 
  minimum disk mass
\begin{equation}
 M_{\rm disk,  min} = \int 2\pi r dr \Sigma_{\rm min}(r)
\end{equation}
  which will bound the general, time dependent disk mass.

\begin{figure}[h!]
\begin{centering}
\includegraphics[width=3.5truein]{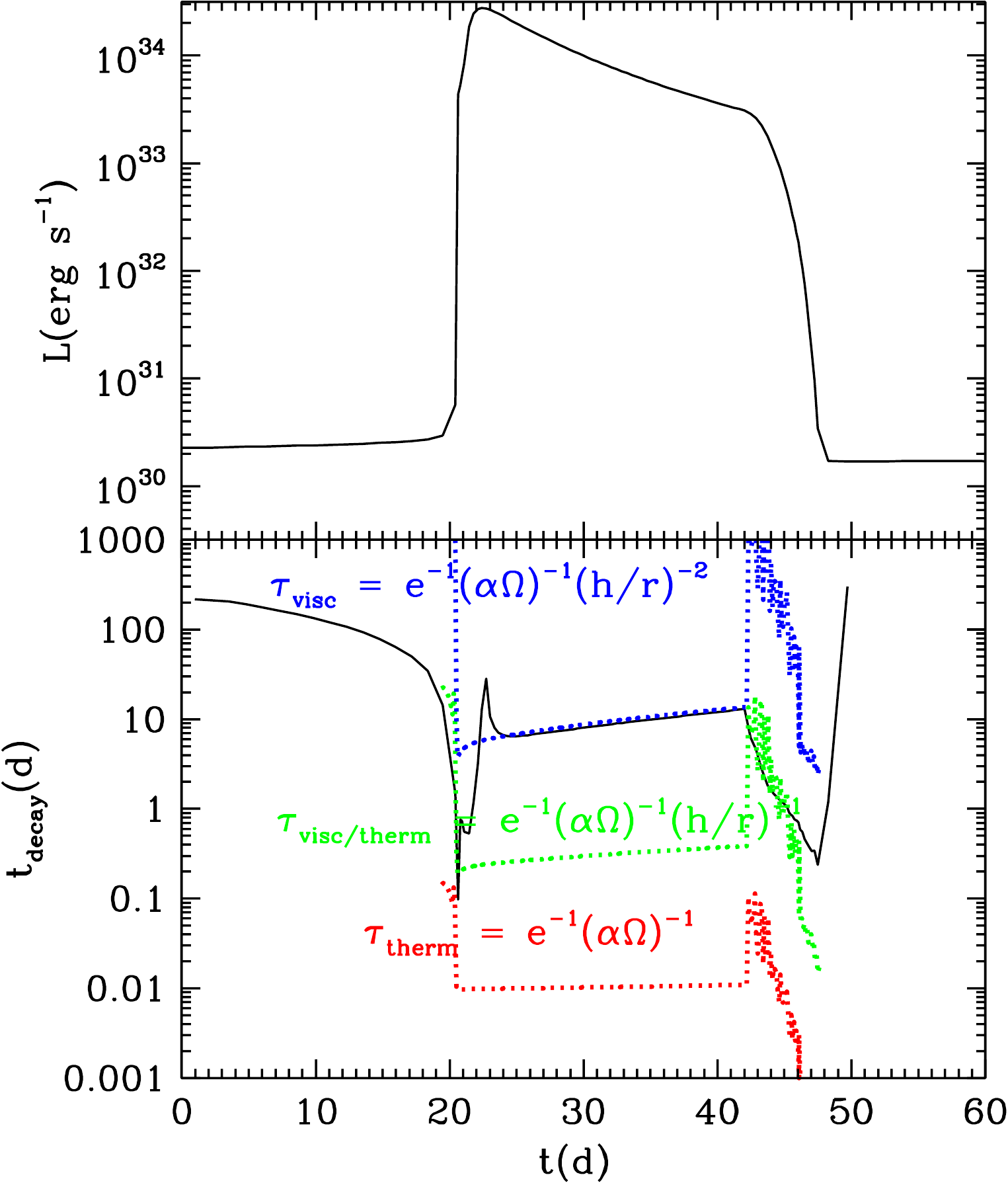}
\caption{The light curve of a superoutburst generated using the ADLC
  model described in \citet{IchikawaOsaki1992} relevant for a SU UMa
  or AM CVn type system (top panel), and the time accretion disk time
  scales (i) $\tau_{\rm visc}$ (blue), (ii) $\tau_{\rm therm}$ (red),
  and (iii) $(\tau_{\rm visc} \tau_{\rm therm})^{1/2}$ (green), all
  evaluated at the outer edge of the hot, actively evolving disk
  (lower panel).  The black curve in the lower panel shows the locally
  defined $e-$folding decay time $|d\log L/dt|^{-1}$ evaluated from
  the top panel light curve.  The disk time scales are normalized by
  $e$ in order to compare to this $e-$folding time.  The viscous time
  sets the slow decay rate and the visco-thermal time sets the
  subsequent faster decay rate, as can be seen by the portions where
  the blue and black curves coincide, and where the green and black
  curves coincide.}
\end{centering}
\label{fig1}
\end{figure}

For normal, `short' outbursts, only a few percent of the stored gas
accretes onto the central object: the thermal time scale of a thin
disk is short compared to the viscous time scale, and the cooling
front that is launched from the outer edge of the disk almost as soon
as the disk enters into outburst traverses the disk and reverts it
back to quiescence.  For disks that have been `filled' to a higher
level with respect to $M_{\rm disk, max}$, the surface density in the
outer disk can significantly exceed the critical surface density
$\Sigma_{\rm min}$.  In order for the cooling front to begin, however,
the outer surface density $\Sigma(r_{\rm outer})$ must drop below
$\Sigma_{\rm min}(r_{\rm outer})$.  Disks in this state generate much
longer outbursts, with slower `viscous' plateaus, because the entire
disk must remain in its high, completely ionized state until enough
mass has been lost onto the WD for the condition $\Sigma(r_{\rm
  outer}) < \Sigma_{\rm min}(r_{\rm outer})$ to be satisfied.
 
Adjacent annuli in the accretion disk are causally connected to each
other via the viscosity.  In any physical problem where processes
operating over a range of time scales are linked, the processes with
the slowest time scales determine the rate of change. In an accretion
disk, the time scales increase with radius, so conditions in the outer
part of the hot, active disk mediate the action.

\begin{figure}[h!]
\begin{centering}
\includegraphics[width=3.3truein]{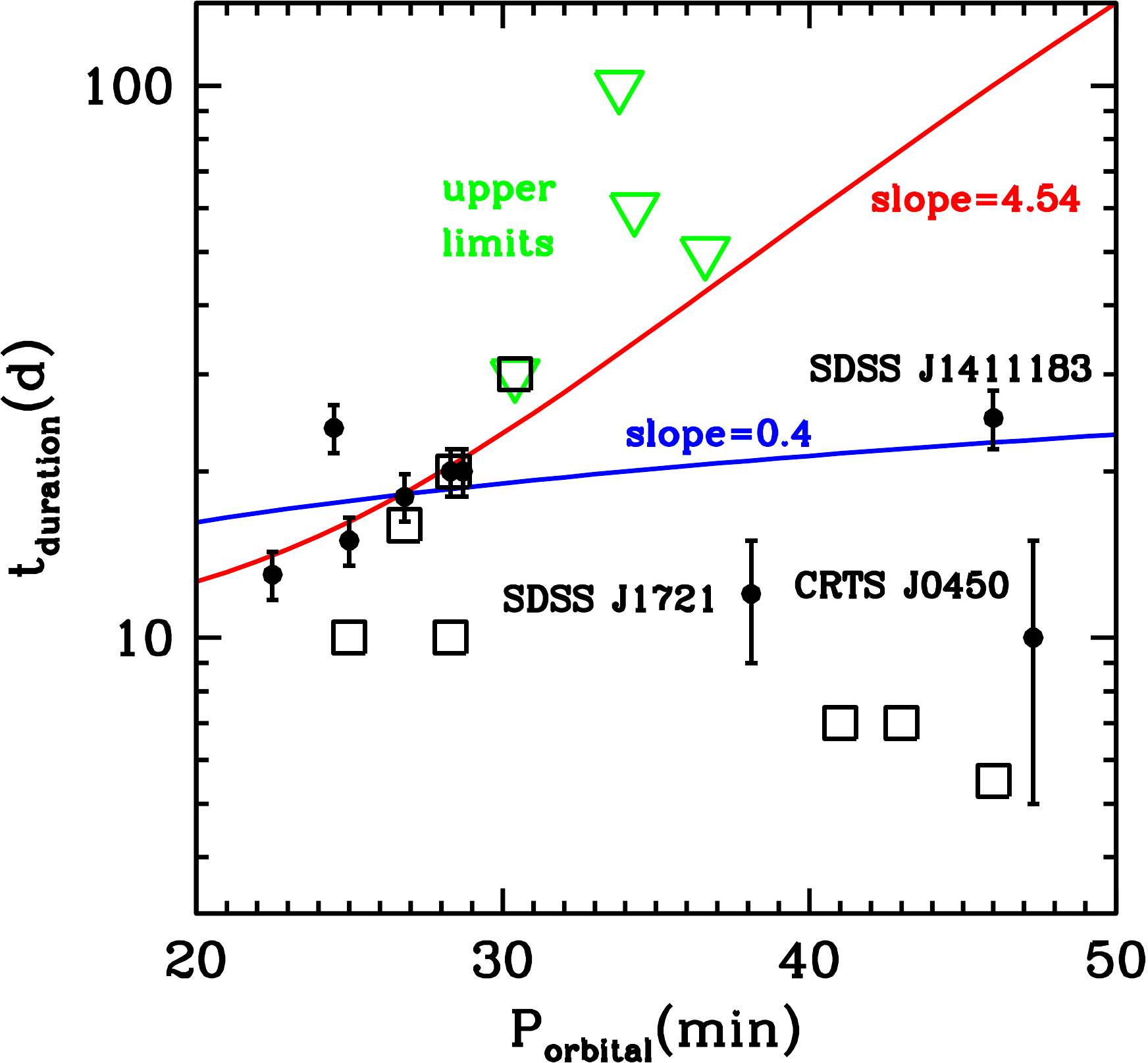}
\caption{ The duration of superoutbursts in AM CVn systems versus
  $P_{\rm orb}$, where we have taken data from \citet{Levitan2015} and
  the literature (see Table \ref{burst-typical}).  The green triangles indicate 75\% of
  the upper limit values from \citet{Levitan2015}, which they combine
  with the data for the six shorter $P_{\rm orb}$ systems to give
  their fit, shown by the red curve.  The blue line indicates the
  slope of the relation expected from the ADLC model
  \citep{CannizzoNelemans2015}.  The open black squares indicate a
  more complete census of superoutburst durations, taken from the
  literature.}
\end{centering}
\label{fig2}
\end{figure}

 Two relevant accretion disk time scales are the viscous time
 $\tau_{\rm visc}$, over which matter is redistributed radially, and
 the thermal time scale $\tau_{\rm therm}$ over which local heating
 and cooling take place.  The slow decay rate within a superoutburst
 provides a direct measure of the viscous time scale in the outer disk
 in the hot state.  The viscous time $\tau_{\rm visc}(r_{\rm
   outer})=[(\alpha_{\rm hot}\Omega_K)^{-1}(h/r)^{-2}]|_{r=r_{\rm
     outer}}$, and the thermal time $\tau_{\rm therm}(r_{\rm
   outer})=[(\alpha_{\rm hot}\Omega_K)^{-1}]|_{r=r_{\rm outer}}$,
 where $\alpha_{\rm hot}$ is the \citet{ShakuraSunyaev1973} viscosity
 parameter in the outbursting disk, $\Omega_K$ is the Keplerian
 angular velocity, and $h/r$ ($\simeq 0.03$) is the disk aspect ratio.
 The subsequent faster decay is governed by $[(\alpha_{\rm
     hot}\Omega_K)^{-1}(h/r)^{-1}]|_{r=r_{\rm outer}}$, the geometric
 mean of $\tau_{\rm visc}$ and $\tau_{\rm therm}$.  This is shown in
 Figure \ref{fig1}.

The decays of DN outbursts at a given orbital period are fairly
uniform and form the basis for the `Bailey' relation, a roughly linear
relation between rate of decay and orbital period
(\citealt{Bailey1975}, \citealt{Warner1995}).  The rise times show a
greater variety, reflecting the fact that the outburst can be
triggered anywhere in the disk (\citealt{Smak1984},
\citealt{Cannizzo98}).  Inside-out bursts tend to produce slow-rise
times, whereas outside-in bursts produce fast rises
\citep{Cannizzo1986}.

\begin{figure}
\begin{centering}
\includegraphics[width=3.5truein]{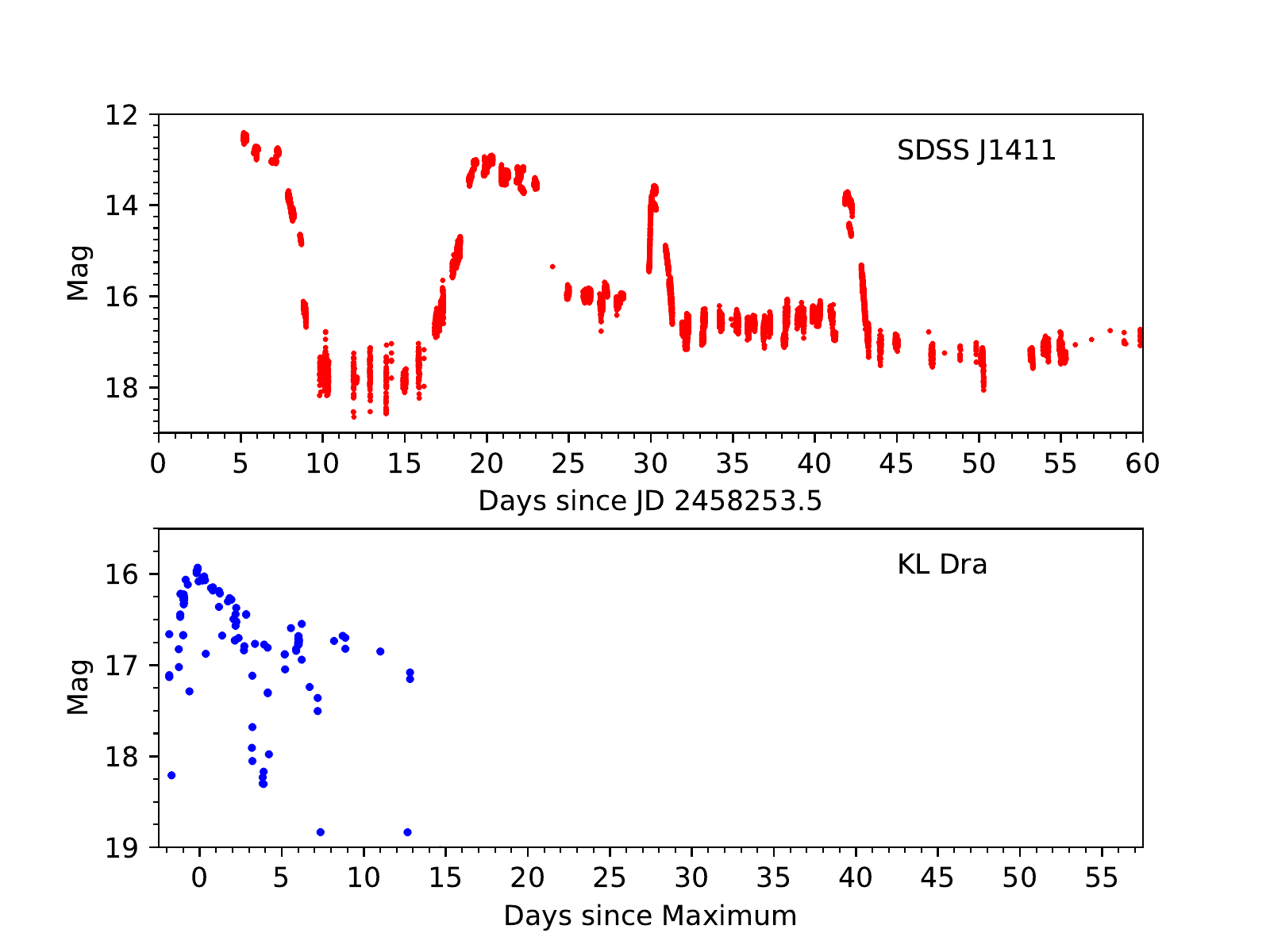}
\caption{A comparison of the May/June 2018 superoutburst of SDSS
  J1411183, taken from the AAVSO (www.aavso.org) data base (top
  panel), with a composite of superoutbursts from KL Dra (bottom
  panel), data from \citet{Ramsay2012}.}
\end{centering}
\label{fig3}
\end{figure}

\section{The Superoutburst Duration}

In this paper we are interested in the duration of the superoutburst,
which we take to be $\tau_{\rm visc}$.  Superoutbursts usually consist
of (i) a rapid rise time, set by $\tau_{\rm therm}$, then (ii) a
period of slow decay, set by $\tau_{\rm visc}$, and finally (iii) a
period of faster decay, set by $(\tau_{\rm visc} \tau_{\rm
  therm})^{1/2}$.  Observationally, the superoutburst duration is
usually contained roughly within the amount of time the transient
spends within $\sim2-3$ magnitudes of peak light.  Some AM CVn systems
have been quoted in the literature as having outbursts lasting months
- e.g., V406 Hya and J0129 \citep{Ramsay2012}, but these include long
outburst tails following the steep decay demarking the end of
superoutburst when the system is significantly brighter than its
nominal quiescence value.  In other words, the relaxation back to the
quiescent state can take much longer than what we define to be $t_{\rm
  dur}$, which is just the duration of the superoutburst itself.
There are probably other factors at work which cause the long periods
of elevated flux post-superoutburst, such as enhanced secondary mass
overflow from the superoutburst, and accretion-induced heating of the
primary.

Figure \ref{fig2} shows superoutburst durations taken from
\citet{Levitan2015} and the literature (Table \ref{burst-typical}).
By including upper limit values as part of their $t_{\rm dur}$
fitting, they obtained a steep relation for $t_{\rm dur}(P_{\rm
  orb})$.  Including only measured values of $t_{\rm dur}$ yields a
much flatter law. We have included the outburst duration from CRTS
J0450 which \citet{Levitan2015} did not include in their fit since
only one outburst was seen.

\begin{table}
\begin{center}
\begin{tabular}{lrrrr}
\hline
Source   & $P_{\rm orb}$           & Duration & Amplitude & Ref \\
             & (mins) & (days)      & (mag) & \\
\hline
PTF1J1919+4815 & 22.5 & 13   & 3.0 & (1)\\
ASASSN-14cc    & 22.5: & 13:   & 3 & (2) \\
CR Boo         & 24.5 & 24 & 3.3 & (3) \\
KL Dra         & 25.0 & 15 & 4.2 & (3)\\
KL Dra         & 25.0 & 10 & 3.3 & (4)\\
PTF1JJ0719+4858      & 26.8 & 16 & 3.5 & (5) \\ 
YZ LMi          & 28.3 & 20 & 2.4 & (3)\\
YZ LMi          & 28.3 & 10 & 2.6 & (4)\\  
CP Eri & 28.4 & 20 & 4.2 & (4)\\
CP Eri & 28.4 & 15 & 4.0 & (3)\\
PTF1 J0943+1029 & 30.4 & 30: & 4 & (6) \\
V406 Hya & 33.8 & $<$50 & 5.2 & (4)\\ 
J0129 & 37: & $<$50 & 4.2 & (4)\\ 
SDSS J1721+2733 & 38.1 & 14: & 4: & (3)\\
ASASSN-14mv & 41: & 7: & 3.5: & (7) \\
ASASSN-14ei & 43: & 7: & 2.5: & (7) \\
SDSSJ1411+4812 & 46 & 25: & 6  & (7, 8) \\
CRTS J0450 & 47.3 & 10 & 5 & (3)\\
\hline
\end{tabular}
\end{center}
\caption{The typical duration, amplitude of those AM CVn systems seen in outburst. 
References: (1) \citet{Levitan2014}, (2) \citet{Kato2015}, (3) \citet{Levitan2015},
(4) \citet{Ramsay2012}, (5) \citet{Levitan2011}, (6)  \citet{Levitan2013}, (7) AAVSO, (8) \citet{RiveraSandoval2019}.}
\label{burst-typical}
\end{table}

\section{Double Superoutbursts and Superoutburst with Dips }

One additional complication is the issue of double superoutbursts and
superoutbursts with dips.  Figure \ref{fig3} compares the 2018
outburst from SDSS J1411+4812 ($P_{\rm orb}=46$ min) taken from AAVSO
observations with that of a composite of superoutbursts from KL Dra
($P_{\rm orb}=25$ min) (taken from \citet{Ramsay2012}).  For SDSS
J1411+4812 the amplitude of the gap between the first and second
portions of the superoutburst is $\sim$5 mag, whereas for KL Dra it is
$\sim$1.5 mag.  Therefore the former may be more aptly described by
`double superoutburst', whereas the latter may be a `dip'.

These phenomena may result from physical effects extrinsic to the ADLC
mechanism, such as irradiation-induced enhanced mass overflow from the
secondary.  Therefore it is not clear whether the entire double
superoutburst should be included in $t_{\rm dur}$ or just the first
one. Given the scatter in Figure \ref{fig2}, however, this issue is
minor.

\section{Conclusions} 

We have investigated the relation between duration of superoutbursts
and orbital period in the AM CVn systems and find a much flatter
relation than inferred by \citet{Levitan2015}.  Their steep dependence
came about primarily because of their inclusion of upper limits of
systems at high $P_{\rm orb}$ in their fit.  By considering solely
measured values of $t_{\rm dur}$ for the high $P_{\rm orb}$ systems
and not using upper limits, we find a much flatter law, consistent
with the flat relation predicted by \citet{CannizzoNelemans2015}.

This work highlights the importance of surveys such as PTF and others
in characterising the long term photometric behaviour of AM CVn
binaries. In particular those with relatively long orbital periods
should be monitored for any new outbursts and high cadence
observations should be obtained to characterise their behaviour,
looking for double superoutbursts, dips and rebrightening events.

\section{acknowledgements}

We thank the AAVSO and its members who contribute their data of
cataclysmic variables and other stars -- they provide an invaluable
resource. Armagh Observatory and Planetarium is core funded through
the Northern Ireland Executive.

\end{document}